\documentclass[preprint2]{aastex}
\slugcomment{submitted to {\it The Astronomical Journal}}
\begin{document}
\title{$uvbyCa$H$\beta$ CCD Photometry of Clusters. V. The Metal-Deficient 
Open Cluster NGC 2243}
\author{Barbara J. Anthony-Twarog\altaffilmark{1}, Jesse Atwell, and Bruce 
A. Twarog\altaffilmark{1}}
\affil{Department of Physics and Astronomy, University of Kansas, Lawrence, 
KS 66045-7582}
\affil{Electronic mail: bjat@ku.edu,jja254@nyu.edu,twarog@ku.edu}
\altaffiltext{1}{Visiting Astronomer, Cerro Tololo Interamerican 
Observatory. CTIO is operated by AURA, Inc.\ under contract to the National 
ScienceFoundation.}

\begin{abstract}
CCD photometry on the intermediate-band $uvbyCa$H$\beta$ system is 
presented for the metal-deficient open cluster, NGC 2243. Restricting the 
data to probable single members of the cluster using the CMD and the 
photometric indices alone generates a sample of 100 stars at the cluster 
turnoff. The average $E(b-y)$ = 0.039 $\pm$0.003 (s.e.m.) or $E(B-V)$ = 
0.055 $\pm$0.004 (s.e.m.), where the errors refer to internal errors alone. 
With this reddening, [Fe/H] is derived from  both $m_1$ and $hk$, using 
$b-y$ and H$\beta$ as the temperature index. The agreement among the four 
approaches is excellent, leading to a final weighted average of [Fe/H] = 
--0.57 $\pm$0.03 (s.e.m.) for the cluster, on a scale where the Hyades has 
[Fe/H] = +0.12. Using a combination of photometric and spectroscopic data, 
27 probable cluster members are identified and used to delineate the red 
giant branch and a well-defined clump at $V$ = 13.70, while eliminating the 
so-called second clump at $V$ = 14.1. Interpolation between isochrones of 
appropriate [Fe/H] leads to an apparent modulus of $(m-M)$ = 13.15 $\pm$0.1 
and an age of 3.8 $\pm$0.2 Gyr. A differential CMD comparison with Berkeley 
29, a cluster with a galactocentric distance almost twice that of NGC 2243, 
constrains Berkeley 29 to be at least as young and as metal-rich as NGC 
2243.
\end{abstract}
\keywords{color-magnitude diagrams --- open clusters and 
associations:individual (NGC 2243, Berkeley 29, Berkeley 54)}

\section{INTRODUCTION}
This is the fifth paper in an extended series detailing the derivation of 
fundamental parameters in star clusters using precise intermediate-band 
photometry to identify probable cluster members and to calculate the 
cluster's reddening, metallicity, distance and age. The initial motivation 
for this study was provided by \citet{tat97}, who used a homogeneous open 
cluster sample to identify structure within the galactic abundance 
gradient. The open clusters appear to populate a bimodal abundance 
distribution with the clusters interior to a galactocentric distance of 
$\sim$10 kpc averaging [Fe/H] $\sim$ 0.0 while those beyond this boundary 
have average [Fe/H] values
of $\sim -0.3$. The dispersion about each mean is less than 0.1 dex and 
implies that within each zone, no statistically significant abundance 
gradient exists, in contrast with the traditional assumption of a linear 
gradient over the entire disk. This structure has since been corroborated 
through the use of Cepheids by \citet{an12,an02,lu03} and, most recently, 
with OB stars \citep{da}. Further evidence for the lack of  a simple linear 
gradient across the entire galactic disk, particularly in the outer 
portion, is found in the spectroscopic data for Berkeley 29 and Saurer 1 
\citep{gc04},
a point we will return to in the last section of this paper. The exact 
origin and reason for the survival of the discontinuity remain unresolved 
issues, though the radial variation in the impact of spiral structure on 
star formation and stellar dynamics appears to be a likely contributor 
\citep{sc01,mi02,lp03}. 

Detailed justifications of the program and the observational approach 
adopted have been given in previous papers in the series 
\citep{AT00a,AT00b,TW03,at04} (hereinafter referred to as Papers I through 
IV) and will not be repeated. Suffice it to say that the reality of the 
galactic features under discussion will remain questionable unless the 
error bars on the data are reduced to a level smaller than the size of the 
effect being evaluated or the size of the sample is statistically enhanced. 
The overall goal of this project is to do both.

An equally important aspect of this research is detailed testing of stellar 
evolution theory as exemplified by comparisons to stellar isochrones based 
upon models derived under a variety of assumptions. The agreement with (or 
the deviation from) the predicted distribution and location of stars within 
the color-magnitude diagram (CMD) has consistently provided a lever for 
adjusting our degree of confidence in the specifics of stellar interiors as 
a function of mass and age. A valuable illustration of this can be found in 
the discussion of the red giant branch distribution (first-ascent and clump 
giants) within NGC 3680, in comparison 
to NGC 752 and IC 4651, clusters of comparable age (Paper IV).

The focus of this paper is the older, metal-deficient open cluster, NGC 
2243. The key role of this cluster in the sequential development of the 
project is as a representative of the moderately metal-poor class of open 
clusters found in the galactic anticenter, a class that contributes 
significantly to the definition of the galactic abundance gradient beyond 
the solar circle. In our context, it typifies the metal-poor population 
found beyond the discontinuity in the galactic abundance gradient at a 
galactocentric radius between 10 and 11 kpc, a population that, to date, 
{\it has no cluster counterpart within the solar circle}, in sharp contrast 
with the field star sample \citep{tat97}.

The early photometric attempts to understand the nature of NGC 2243 from 
CMD and color-color diagrams were based upon a mixture of photoelectric and 
photographic data \citep{ha, vb}. While agreeing that the cluster was 
metal-deficient relative to the sun, the disagreement over the exact value 
of the metallicity, and most of the other key cluster parameters, was 
significant. There have since been three CCD-based studies, 
\citet{be,bo,ka} (hereinafter referred to as BE, BO, and KA, respectively). 
The first two used $BV$ photometry, while the last used $VI$. We will defer 
discussion of their conclusions, as well as those of other investigators, 
regarding the cluster parameters to the sections of the paper where they 
are derived within the present analysis.

Section 2 contains the details of the $uvbyCa$H$\beta$ CCD observations, 
their reduction and transformation to the standard system, and a search for 
photometrically anomalous stars. In Sec. 3 we discuss the CMD and begin the 
process of identifying the sample of probable cluster members. Sec. 4 
contains the derivation of the fundamental cluster parameters of reddening 
and metallicity. In Sec. 5, these are combined with broad-band data to 
derive the distance and age through comparisons with theoretical isochrones 
and to allow a differential comparison with the most distant open cluster, 
Berkeley 29 (hereinafter Ber 29). Sec. 6 summarizes our conclusions 
regarding NGC 2243, Ber 29 and their role with respect to the abundance 
gradient of the galactic disk.

\section{The Data}
\subsection{Observations: CCD $uvbyCa$H$\beta$}
The new photometric data for NGC 2243 were obtained using the 
Cassegrain-focus CCD imager on the National Optical Astronomy Observatory's 
0.9-m telescope at Cerro Tololo Interamerican Observatory. We used a 
Tektronix 2048 by 2048 detector at the $f/13.5$ focus of the telescope, 
with CTIO's $4'' \times 4''$ $uvby$ filters and our own $3'' \times 3''$ 
H$\beta$ and $Ca$ filters.  The field size is $13.5'$ on a side.  Frames of 
all seven filters were obtained in both November 2000 and January 2002, and 
were pre-processed (bias subtraction, trimming and flattening) through IRAF 
routines at the telescope.  As over 110 frames were ultimately used, we have
elected not to present a detailed exposure log but it is useful to note
that we were able to combine information from 13 to 17 frames in each color
(24 for the $u$ filter), with the following exposure totals
for $y$, $b$, $v$, $u$, $Ca$, $\beta w$, and $\beta n$ : 9, 24, 89, 293,
93, 33 and 176 minutes.

\subsection{Reduction and Transformation}
Previous papers in this series describe the procedures used to produce high 
precision, accurately calibrated photometry from CCD data. The ALLSTAR 
routine within the IRAF DAOPHOT package was used to obtain complete sets of 
profile-fit magnitudes for all stars on every program frame.  Paper I, in 
particular, provides a comprehensive description of the steps used to 
produce
average instrumental magnitudes and indices of high precision.
 
Regardless of the internal precision achieved by averaging large numbers of 
frames, the accuracy of the photometric calibration is limited by other 
factors, including the breadth of parameter space covered by standard stars 
and observational conditions.  
Our approach for these steps is described extensively in Paper IV; since 
NGC 3680 and NGC 2243 were both observed on photometric nights in January 
2002, the backbone of the calibration described for NGC 3680 is also used 
to calibrate NGC 2243.  

Briefly, standard stars in the field and in clusters, as well as uncrowded 
stars in program  clusters, are observed on photometric nights and reduced 
using a consistent aperture measurement strategy.  This permits the 
standardization of the aperture photometry in the program cluster;  the 
calibration is extended to the more precise and generally deeper 
profile-fit magnitudes and indices by determining the average offset 
between the aperture photometry and profile-fit indices.

As described in Paper IV, the calibration equations developed for 
photometric nights in January 2002 carry standard errors of the mean for 
the zero point of 0.007 for $\beta$, 0.005 for $hk$, 0.006 for $V$ and 
$b-y$, 0.009 and 0.010 for dwarf stars' $m_1$ and $c_1$ indices and, with 
somewhat higher uncertainties, 0.010 and 0.017, for the cool giant stars' 
$m_1$ and $c_1$ indices.  

Establishing the link between the high precision set of indices based on 
profile-fit photometry and the calibrated aperture photometry introduces a 
modest increment to the zero-point uncertainties for each calibrated index. 
 The mean differences between these sets of indices were established with 
standard errors of the mean for $V$, $b-y$, $m_1$, $c_1$, $hk$ and $\beta$ 
of 0.002, 0.002, 0.003, 0.004, 0.003 and 0.012.

Final photometric values are found in Table 1, where the primary 
identification and coordinate description follow the WEBDA database 
conventions for this cluster as of July 2004. Stars not found in the WEBDA 
database have identification numbers above 10000. Following each of the six 
photometric indices are the standard errors of the mean for each index and 
a summary of the number of frames in each of the seven bandpasses. 
For stars with $V$ $\lesssim$ 16.5 and $b-y$ $\geq$ 0.45, the $m_1$ and 
$c_1$ instrumental indices 
were transformed using the calibration relations appropriate for giant 
stars.
Figs. 1 and 2 show the average standard error of the mean for each index as 
a function of $V$ as well as
the dispersion about the average value. 

\subsection{Potential Variables}
The photometry discussed above was obtained to reliably define the cluster 
in intermediate-band CMD and color-color diagrams so, while over 100 frames 
have been analyzed, the distribution is approximately 15 frames per filter 
at the bright end and less than half that among the fainter stars. Despite 
the lack of optimization for detecting variable stars, it should be 
possible to identify at least some candidates with modest to large 
amplitude variations over a range of timescales. An example of the 
appropriate methodology for identifying cluster variables may be found in 
the study of NGC 2243 by KA; this paper and its six variable stars will 
also provide a test of our success in finding potential variables.

The obvious criterion for discovering variables is a large photometric 
scatter in the indices for a star relative to what is expected at a given 
magnitude. The points that deviate from the well-defined trends in Figs. 1 
and 2 should generate a first cut for our sample. However, deviant points 
could exist in Figs. 1 and 2 for a variety of reasons, most of which have 
nothing to do with variability. Since we are plotting the standard error of 
the mean, two stars at the same magnitude level with identical 
frame-to-frame photometric errors will populate different locations in the 
figure because one star only appears on a small subset of the frames. The 
lack of a complete set of magnitudes is a common occurrence for stars in 
the outer edges of the cluster field, as the cluster position on the CCD 
chip is shifted from night to night and run to run to avoid placing the 
same stars on the same bad columns and pixels, and to extend to field 
coverage beyond the minimum allowed by one CCD frame. Additionally, a 
paucity of measures for a star in any region of the cluster may be a sign 
of image overlap/confusion, resulting in poor to inadequate photometry. 
Finally, while the frames have been processed using a 
positionally-dependent point-spread-function, the success of such 
modifications may decline as one nears the edge of the chip and the 
photometric scatter in any filter may be expected to increase in the outer 
zone of the field, irrespective of the number of frames included in the 
average.

To tag possible variables, the stars that deviated the most from the mean 
relations for $V$ and $b-y$ in Fig. 1 for $V$ brighter than 18.05 were 
identified and the errors renormalized to a uniform number of frames; the 
reason for the artificial choice of the lower bound will become apparent 
below. If the renormalization placed the star near the standard relation, 
it was excluded. Stars located more than 600 pixels ($\sim$ $6.1'$) from 
the cluster center were excluded because beyond that distance from the
center of the field, photometric errors traceable to spatial variations 
in the psf begin to make selection of variables difficult. 
Stars appearing on 
fewer than 2/3 of the expected frames at a given magnitude were checked for 
possible contamination/confusion with a neighbor and eliminated if a 
companion was found. Finally, the individual magnitude errors were checked 
and a star retained if it exhibited large scatter in $y$ and at least one 
additional filter among $b$, $v$, and $Ca$, and showed a larger than 
expected error in H$\beta$. The final result was the identification of the 
nine most likely variables within the sample: 372, 1419, 1463, 1558, 1583, 
1728, 1746, 2363,  and 2445. These stars are noted by filled symbols in 
Figs. 1 and 2. 

Of the nine stars, five (372, 1558, 1583, 1728, and 2363) have been 
identified as variables by KA. The $V$ limit for the search was extended to 
reach the complete sample of KA with 1583 at $V$ = 18.05. The sixth known 
variable (2853) does not show significant scatter in any index. 

 \subsection{Comparison to Previous Photometry}
Only a handful of stars, all red giants, have been observed 
photoelectrically in $vby$ by \citet{ri}, but there have been three 
broad-band CCD surveys of NGC 2243 that allow us to, at minimum, check the 
$V$ magnitude system and identify potential photometric deviants. The two 
earlier studies in $BV$, BO and BE, both include two overlapping fields 
near the cluster center observed with small-format CCD's. The former data 
were standardized using older photoelectric observations in the cluster 
\citep{vb} while the latter were tied to a mixture of field standards from 
\citet{la3} and \citet{gra}. The $VI$ data of KA cover an area comparable 
to the current study and were standardized to the system of \citet{la2}.

As in the previous papers in the series, we will compare the residuals in 
the photometry for the entire sample and for the brighter stars alone since 
the reliability of the latter sample is more relevant to the determination 
of the cluster parameters. Our interest in the residuals is twofold: to 
measure the photometric uncertainties in comparison with what is expected 
from the internal errors and to identify stars that exhibit significant 
deviations from one study to another. The deviants will be a mixture of 
misidentifications, bad photometry, and, of particular value, long-term 
variables or longer-period eclipsing binaries that will not be readily 
exposed by the variable star searches done to date. Data for the published 
surveys have been taken from the WEBDA Cluster Data Base.

For each survey, the residuals in $V$, in the sense (Table 1 - Ref), were 
calculated for all stars in common to both and for all stars brighter than 
$V$ = 17.5. For the complete samples, all stars with residuals more than 
0.1 mag above or below the survey mean were excluded from the analysis and 
the mean and dispersion among the residuals rederived. For the brighter 
sample, the procedure was repeated, but the exclusion limit was reduced to 
0.075 mag above or below the mean offset. The residuals in all comparisons 
were checked for color terms and a magnitude dependence. Two significant 
trends were identified. For both the brighter stars and the complete sample 
comparisons, a magnitude dependence is present in $V$ for the data of KA. 
The fact that the other two surveys do not exhibit this problem implies 
that the error lies with the photometry of KA. For the $V$ data of BO, a 
modest color term was noted in comparisons with the three other data sets.

The $V$ data of KA  and BO were transformed to the system of Table 1 using 
the linear relations noted for each in Table 2 for the brighter stars and 
the residuals recomputed. The results for the revised data, as well as the 
other comparisons, may be found in Table 2.

For BO, there are 402 stars that overlap with Table 1; of these 21 have 
residuals more than 0.10 magnitudes away from the mean. For the remaining 
381 stars, the average offset is --0.011 $\pm$ 0.029; inclusion of the 
color term lowers the dispersion to 0.028. Turning to the 215 stars 
brighter than $V$ = 17.5, 206 have residuals within 0.075 of the cluster 
mean. Of the nine deviants, stars 1761 and 1793 appear to be switched in 
the WEBDA data base, while 1728 and 2363 are known variables from KA. The 
remaining deviants are 1012, 1419, 1707, 1995, and 2414. Star 1419 has been 
noted in the previous section as having larger photometric scatter than 
expected for its magnitude. Star 2414 is clearly incorrect in BO, being 
significantly redder than the published photographic and CCD data for the 
star, though the error in BO may be a product of severe crowding or 
misidentification. The photometry in Table 1 for 1012, 1707, and 1995 is in 
excellent agreement with both BE and KA while 1419 disagrees with all three 
studies. For the 206 non-deviants, the mean offset is --0.014 $\pm$ 0.021. 
Inclusion of the color term changes this to 0.000 $\pm$ 0.018.

For BE, there are 396 stars that overlap; 379 have residuals within 0.10 
mag of the mean. For the non-deviant sample, the average offset is +0.007 
$\pm$ 0.028. Restricting the sample to the 225 brighter stars generates 5 
deviants. The remaining 220 stars have a mean offset of +0.004 $\pm$ 0.019. 
Of the five deviants, 3633 is misidentified in the database and there are 
two stars listed as 2236; the brighter star is the correct one. This leaves 
only the already-noted 1419, 1931, and 1510. The photometry for 1931 in 
Table 1 agrees with the results found in BO and KA.

For KA, using the entire sample, 930 stars overlap, of which 35 are 
excluded. The mean offset for the remaining 895 stars is +0.019 $\pm$ 
0.034. After application of the magnitude relation, only 29 stars are 
excluded and, by definition, the mean offset becomes 0.000 with a 
dispersion of 0.031. For the brighter sample, 414 stars overlap with 7 
stars exhibiting large deviations. Without the magnitude adjustment, the 
mean offset is +0.034 $\pm$ 0.021; with the magnitude dependence removed, 
only 5 stars are tagged as deviants and the offset becomes 0.000 $\pm$ 
0.018. The five anomalous stars include the known variable 1728, two stars 
identified in the previous section as possible variables, 1419 and 2445, 
1725, which may be affected by crowding with 1729, a star that barely 
missed being classified as a deviant, and 1510, noted above.

\section{The Color-Magnitude Diagram: Thinning the Herd}

Because of the lack of membership information for any of the stars except 
the few with radial velocities and the significant distance modulus of the 
cluster, NGC 2243 will be treated as a program cluster analogous to NGC 
6253 (Paper III) rather than NGC 3680 (Paper IV).

The CMD for all stars with at least 2 observations each in $b$ and $y$ is 
presented in Fig. 3. Open circles are the stars with standard errors in the 
mean $\leq$ 0.010 mag for $b-y$. Though the superposition of so many points 
makes it difficult to resolve, the majority of stars brighter than $V$ 
$\sim$ 17.0 meet the rather stringent limit on the precision in $b-y$. The 
morphology of the CMD is similar to that found in previous studies. The 
turnoff region is well defined to $V$ = 16 where an obvious change in the 
distribution of stars in color occurs. This dramatic broadening is the 
product of a main sequence contaminated in large part by a significant 
population of binaries; the sloping binary sequence merges with the 
vertical turnoff to generate the rather pinched off appearance near the 
upper turnoff. 

The exact location of the subgiant branch is difficult to define between 
$b-y$ = 0.45 and 0.6 due to the field star contamination coming from stars 
in the outer region of the CCD field, but the first-ascent giant branch is 
discernible from the red limit of the distribution of stars above $V$ = 15. 
The question of the correct location of the red giant clump reasserts 
itself here. The two options are near $V$ = 13.7, $b-y$ = 0.60 and $V$ = 
14.1, $b-y$ = 0.53. How these two groups of stars are related, if at all, 
is a key question we will address later. Finally, five obvious candidates 
for blue straggler members can be seen in isolation well blueward of the 
turnoff and brighter than $V$ = 15.5.

\subsection{Thinning the Herd: The Radial Distribution}
A straightforward means of enhancing the probability of cluster membership 
is to select stars closest to the cluster core where the ratio of members 
to field stars is the highest. The stellar radial distribution has been 
investigated by both \citet{vb} using photographic plates and KA using CCD 
data; the surveys reach $V$ $\sim$ 21 and $I$ = 20.4, respectively. Taking 
into account the $(V-I)$ color of the main sequence, the CCD study has a 
$V$ limit of $\sim$21.8. Both studies conclude that while the stellar 
distribution with radius flattens beyond about $4'$, the cluster can be 
traced to at least $6'$ away from the cluster center. The form of the 
cluster CMD is readily identified in the sample of stars beyond $5'$ in 
Fig. 6 of KA. It should be noted, however, that the degree of concentration 
changes as the sample expands to include fainter stars. To illustrate this 
point, we superpose the radial distribution from our CCD data, complete to 
$V$ = 18.0 (circles), on top of those of \citet{vb} (squares) and KA 
(triangles) in Fig. 4. The analysis of the our sample was done by first 
identifying the most probable center of the cluster using the same approach 
as in Paper III, rather than simply eyeballing the location. The surface 
densities have been normalized to ensure that the 
extended flat regions of the distribution 
between 300\arcsec\ and 400\arcsec\ 
have the same mean level as that of KA. 
Error bars are generated only for the inner data points for clarity since 
the error bars are smaller than the points for the the outer regions. 

The higher degree of concentration for the brighter sample compared to 
\citet{vb} is not unexpected given the three magnitude difference in the 
limiting magnitude of the counts, the increasing likelihood of field star 
contamination at fainter $V$, and a possible weakening of the present-day 
luminosity function toward lower mass stars due to mass segregation and 
cluster evaporation. The change in profile between \citet{vb} and KA is 
somewhat surprising given the more modest change in the limiting magnitude. 
This may indicate that the luminosity function of the cluster undergoes a 
decline beyond $V$ = 21 since the mass difference in the limiting 
magnitudes of the two surveys is too small to generate large mass 
segregation effects.

To maximize the membership probability, our spatial cut will restrict the 
sample to stars within 200 pixels of the cluster center, just over $120"$ 
in Fig. 4. The improvement in delineating the CMD is easily seen in Fig. 5, 
which has the same symbols as Fig. 3, but includes only the cluster core. 
Note that four of the five probable blue stragglers are retained, the 
subgiant and giant branches are readily identifiable, and most of the stars 
that scattered redward of the main sequence are eliminated.

\subsection{Thinning the Herd: CMD Deviants}
For purposes of optimizing the derivation of the cluster reddening and 
metallicity, our interest lies in using only single stars that evolve along 
a traditional evolutionary track and that have indices in a color range 
where the intrinsic photometric relations are well defined. With this in 
mind, we further reduce the sample by including only stars with errors in 
$b-y$ below 0.010, in the magnitude range from $V$ = 15.5 to 17.5 and the 
color range from $b-y$ = 0.28 to 0.42, leaving 119 stars.

Given the high precision of the $b-y$ indices and the expected dominance of 
the cluster sample over the field stars, it is probable that the majority 
of stars at the cluster turnoff are members, though not necessarily single 
stars. However, unlike IC 4651 in Paper I and NGC 3680 in Paper IV, the 
larger distance of NGC 2243 places the majority of the stars under 
discussion within the vertical portion of the turnoff. Separation of the 
stars into two parallel sequences, one single and one composed of binaries, 
becomes a challenge because the single stars evolving away from the main 
sequence produce an evolutionary track that curves toward and crosses the 
rich binary sequence composed of unevolved pairs. 

In previous papers, we have enhanced our ability to delineate the single 
and binary sequences through the use of $u-y$, the color index with the 
largest baseline in wavelength and the greatest sensitivity to temperature 
change. The one weakness of this approach is with the evolved stars at the 
turnoff. As explained in Paper IV, the $u-y$ index for binaries composed of 
unevolved stars follows a simple relation between effective temperature and 
color. For single stars in the vertical band at the turnoff, evolution off 
the main sequence alters the energy distribution by decreasing the relative 
contribution of the ultraviolet region blueward of the Balmer 
discontinuity. Thus, $c_1$ increases as $M_V$ decreases at a given 
temperature. The declining contribution of the $u$ filter leads to a redder 
$u-y$ index due to surface gravity effects rather than temperature, making 
the more evolved stars appear redder than less evolved stars at the same 
temperature. The result is that the binary sequence will cross the vertical 
turnoff at a redder and fainter location than in $b-y$, thereby lessening 
its value for a cluster sample such as ours in NGC 2243. 

An intermediate alternative is to use the $v-y$ index championed by 
\citet{me00}. The greater baseline gives the index greater temperature 
sensitivity than $b-y$ but, since $v$ is dominated by metallicity effects 
rather than surface gravity and all the stars within the cluster supposedly 
have the same [Fe/H], the merger of the two sequences should be less of an 
issue.  For the 119 stars remaining after our previous cuts, the $V, v-y$ 
diagram is illustrated in Fig. 6. For stars redder than $v-y$ = 0.8, the 
main sequence is displayed as a band 0.75 mag wide in the vertical 
direction, exactly as expected for a range of pairs with the upper bound 
generated by two identical stars. For stars redder than $v-y$ = 0.82 we 
have split this band down the middle and classified all the stars in the 
upper half (filled circles) as potential binaries and/or nonmember 
interlopers. To test whether this separation makes any sense, we have 
plotted the same stars with the same symbols in both a $V, b-y$ and a $V, 
u-y$ diagram. Since we draw the same conclusion from both, we will only 
show the former diagram in Fig. 7. The majority of the stars identified as 
potential binaries in Fig. 6 fall redward of the primary sequence on Fig. 
7, as expected if they were, in fact, binaries or nonmembers. Given the 
photometric uncertainty and our restricted selection in Fig. 6, 
the relatively small number of non-binary stars deviating from the blue edge in
Fig. 7 is reassuring. 
Only two of the many stars that deviate 
to the red in $u-y$ remained classified as single stars in Fig. 6. As 
stated in past papers, exclusion of some single stars from the final sample 
is of no consequence if isolating potentially anomalous stars/photometry 
allows us to exclude a reasonable portion of the data that could distort 
the final parameters. Removing the filled circles leaves us with 100 
probable, single-star members.

\section{Fundamental Properties: Reddening and Metallicity}
\subsection{Reddening}

All 100 stars have multiple measures in every filter. The average standard 
errors of the mean for the various indices are 0.0074, 0.0094, 0.0083, and 
0.0062 for $b-y, m_1, c_1$, and H$\beta$, respectively. For H$\beta$, only 
two stars have standard errors above 0.010 mag. None of the remaining stars 
has been excluded because of larger than average errors.

As discussed in Paper I, derivation of the reddening from intermediate-band 
photometry is a straightforward, iterative process given reliable estimates 
of H$\beta$ for each star. The primary decision is the choice of the 
standard relation for H$\beta$ versus $b-y$ and the adjustments required to 
correct for metallicity and evolutionary state. The two most commonly used 
relations are those of \citet{ols88} and \citet{ni88}. As found in previous 
papers for IC 4651, NGC 6253, and NGC 3680, both produce very similar if 
not identical results. (A correction regarding the citations should be 
mentioned for Papers III and IV. Though we have used the same procedures 
for all open clusters and no changes are required in the numbers or the 
conclusions, the second standard relation cited in Papers III and IV is 
\citet{sn89}; in both cases, this should read \citet{ni88}.)

Processing the indices for the 100 stars through both relations generates 
$E(b-y)$ = 0.042 $\pm$0.022 (s.d.) with \citet{ols88} and $E(b-y)$ = 0.037 
$\pm$0.017 (s.d.) with \citet{ni88}. We will take the weighted average of 
the two and use $E(b-y)$ = 0.039 $\pm$0.003 (s.e.m.) or $E(B-V)$ = 0.055 
$\pm$0.004 (s.e.m.) in the analyses that follow.

The possibility of variable reddening across the extent of the cluster was
considered and addressed by examining a sample of $\sim500$ stars between
$V = 16$ and 18 with colors typical of the turnoff and errors in $(b-y)$ and
H$\beta$ below 0.02 mag.  The color excess with
respect to a mean relation between $(b-y)$ and H$\beta$ was examined for trends
with respect to coordinate position.  No significant trend is discernible,
presumably since any plausible range in reddening value would
not be very large compared to realistic photometric errors in the colors.

\subsection{Metallicity from $m_1$} 
Given the reddening of $E(b-y)$ = 0.039, the derivation of [Fe/H] from the 
$m_1$ index is as follows. The $m_1$ index for a star is compared to the 
standard relation at the same color and the difference between them, 
adjusted for possible evolutionary effects, is a measure of the relative 
metallicity. Though the comparison of $m_1$ is often done using $b-y$ as 
the reference color because it is simpler to observe, the preferred 
reference index is H$\beta$ due to its insensitivity to both reddening and 
metallicity. Changing the metallicity of a star will shift its position in 
the $m_1$ - $(b-y)$ diagram diagonally, while moving it solely in the 
vertical direction in $m_1$ - H$\beta$. Moreover, reddening errors do not 
lead to correlated errors in both $m_1$ and H$\beta$. 

In past papers, we have derived the metallicity using $b-y$ and H$\beta$ as 
the defining temperature index for $m_1$ with, on average, no statistically 
significant difference in the outcome.  Since the publication of Paper IV, 
alternative [Fe/H] calibrations based upon $b-y$ and $m_1$ have been 
derived by \citet{no04} for F stars and cooler, calibrations that
make use of the reddening-corrected indices rather than differentials 
compared to a
standard relation. This approach was first used successfully by 
\citet{sn89}, but the
primary focus of their work was on metal-deficient dwarfs and concerns 
about the application of the function to solar-metallicity dwarfs limited 
its adoption for disk stars. 
These concerns proved valid for the metallicity calibration for cooler 
dwarfs where [Fe/H] was systematically underestimated at the metal-rich end 
of the scale \citep{tw02}. The more extensive recalibrations for F dwarfs 
and cooler by \citet{no04} are readily applicable to solar and higher 
metallicity dwarfs at all colors and eliminate the concerns regarding the 
original functions of \citet{sn89}. We will derive [Fe/H] without reference 
to H$\beta$ using the \citet{no04} relation and from 
$\delta$$m_1$(H$\beta$) as in past papers, supplying an independent check 
of the calibration relations.
 
After correcting each star for the effect of $E(b-y)$ = 0.039, the mean 
[Fe/H] using the 
F-star relation of \citet{no04} is --0.580 $\pm$ 0.028 (s.e.m.). In 
contrast, after correcting each star for the effect of $E(b-y)$ = 0.039 and 
deriving the differential in $m_1$ relative to the standard relation at the 
observed H$\beta$,  the average $\delta$$m_1$ for 100 stars is +0.056 
$\pm$0.002 (s.e.m.), which translates into [Fe/H] = --0.534 $\pm$0.023 
(s.e.m.) for the calibration as defined in \citet{ni88} and adopted in 
previous papers. Note that the error for an individual estimate from the 
first approach is larger than the $\delta$$m_1$(H$\beta$) technique because 
of the enhanced sensitivity of the non-linear terms in the first [Fe/H] 
calibration to errors in both $b-y$ and $m_1$. The agreement between the 
averages is quite good, but  improves even more when one takes the 
zero-point of the [Fe/H] scale into consideration. 

The zero-point of the H$\beta$ metallicity calibration has been fixed to 
match the adopted 
value for the Hyades of [Fe/H] = +0.12, {\it i.e.}, if one processes the 
data for the Hyades or the 
standard relation through the [Fe/H] calibration, one is guaranteed to 
obtain 
[Fe/H] = +0.12 for any star with $\delta$$m_1$ = 0.000. If the standard 
relation or 
the observed data for the Hyades are processed through the \citet{no04} 
relation, 
at the cooler end of the scale beyond $b-y$ = 0.32, one obtains [Fe/H] 
between = +0.12 
and +0.16. As $b-y$ decreases, [Fe/H] declines steadily, reach a minimum 
near +0.03 
near the hotter end of the scale ($b-y$ = 0.23). For the stars in the color 
range of interest 
for NGC 2243,  [Fe/H] for the Hyades is systematically underestimated by 
0.05 dex 
relative to the adopted value. Thus, for consistency on the scales,  the 
[Fe/H] estimate
based upon $b-y$ and $m_1$ should be raised to --0.53, essentially 
identical 
to the H$\beta$-based relation.

The primary weakness of metallicity determination with intermediate-band 
filters is the sensitivity of [Fe/H] to small changes in $m_1$; the typical 
slope of the [Fe/H]/$\delta$$m_1$ relation is 12.5. Even with highly 
reliable photometry,
{\it e.g.}, $m_1$ accurate to $\pm$0.015 for a faint star, the uncertainty 
in [Fe/H] for an individual star is $\pm$0.19 dex from the scatter in $m_1$ 
alone. When potential photometric scatter in H$\beta$ and $b-y$ are 
included, errors at the level of $\pm$0.25 dex are common, becoming even 
larger for polynomial functions of the type discussed above. As noted in 
previous papers in this series, the success of the adopted technique 
depends upon both high internal accuracy and a large enough sample to bring 
the standard error of the mean for a cluster down to statistically useful 
levels, {\it i.e.}, below $\pm$0.10 dex. Likewise, because of the size of 
the sample, we can also minimize the impact of individual points such as 
binaries and/or the remaining nonmembers, though they will clearly add to 
the dispersion.

\subsection{Metallicity from $hk$}
We now turn to the alternative avenue for metallicity estimation, the $hk$ 
index. The $hk$ index is based upon the addition of the $Ca$ filter to the 
traditional Str\"{o}mgren filter set, where the $Ca$ filter is designed to 
measure the bandpass that includes the H and K lines of Ca II. The design 
and development of the $Caby$ system have been laid out in a 
series of papers discussing the primary standards \citep{att91}, an 
extensive 
catalog of field star observations \citep{tat95}, and calibrations for both 
red giants \citep{att98} and metal-deficient dwarfs \citep{att00}. Though 
the system was optimally designed to work on metal-poor stars and most of 
its applications have focused on these stars \citep{atc95,bd96,att00}, 
early indications that the system retained its metallicity sensitivity for 
metal-rich F dwarfs have been confirmed by observation of the Hyades and 
analysis of nearby field stars \citep{at02}. What makes the $hk$ index, 
defined as $(Ca-b)-(b-y)$, so useful for dwarfs, even at the metal-rich end 
of the scale, is that it has half the sensitivity of $m_1$ to reddening and 
approximately twice the sensitivity to metallicity changes. The metallicity 
calibration for F stars derived in \citet{at02} used $\delta hk$ defined 
relative to $b-y$ as the temperature index. To minimize the impact of 
reddening on metallicity, this calibration was redone in Paper III using 
H$\beta$ as the primary temperature index, leading to the preliminary 
relation

\medskip
\centerline{[Fe/H]$ = -3.51 \delta hk(\beta) + 0.12$}
\smallskip
\noindent
with a dispersion of only $\pm$0.09 dex about the mean relation. Though the 
derived zero-point of the relation was found to be +0.07, it was adjusted 
to guarantee a Hyades value of +0.12, the same zero-point used for the 
$m_1$ calibration. Because of the expanded sample of stars with 
spectroscopic abundances and the revised $m_1$ calibrations for F and G 
dwarfs by \citet{no04}, a revised and expanded calibration of the 
$\delta$$hk$ indices, based on both $b-y$ and H$\beta$ is underway 
\citep{act}.  Modest changes have been generated in the color dependence of 
the [Fe/H] slope for $\delta$$hk(b-y)$, with even smaller adjustments to 
the H$\beta$-dependent relation. To ensure that the metallicities based 
upon $m_1$ and $hk$ are on the same internal system, we have generated 
[Fe/H] from the unmodified $m_1$,$(b-y)$  function of \citet{no04} for all 
dwarfs with $hk$ indices and derived linear relations between [Fe/H] and 
$\delta$$hk(b-y)$ and $\delta$$hk($H$\beta)$ for three different color 
ranges among the F-stars. 

Applying these modified metallicity calibrations to the $hk$ data for 100 
stars
in NGC 2243, the resulting [Fe/H] values for $hk$ relative to $b-y$ and 
H$\beta$ are [Fe/H] = --0.643 $\pm$0.018 (s.e.m.) and --0.643 $\pm$0.011 
(s.e.m.), respectively. It should be emphasized that the exact agreement 
between the two is purely fortuitous given the color sensitivity of the 
$b-y$ calibration, as evidenced by the larger dispersion for the 
$(b-y)$-based [Fe/H] determination. The dramatic decrease in the scatter 
with the
H$\beta$ relation is real and an indicator of the value of coupling the 
increased
metallicity sensitivity of the $hk$ index with the reddening and 
metallicity-independent 
H$\beta$ index and the minimal temperature-dependence of the [Fe/H] 
calibration
based upon H$\beta$. If we repeat the adjustment derived above to ensure 
that the
calibrations are zeroed to the same scale, the [Fe/H] from the $hk$ index 
becomes
--0.59. Note that this is identical to the average we would have obtained  
from the 
calibrations used in the previous papers, though the individual 
$(b-y)$-based and 
H$\beta$-based averages would not have been identical. The unweighted 
average of the four determinations is [Fe/H] = --0.56 $\pm$ 0.03, while 
inclusion of a weight based upon the inverse of the standard error of the 
mean lowers the average to [Fe/H] = --0.57. Given that the photometric 
zero-points for the $m_1$ and $hk$ photometry were derived independently of 
each other and that the reddening impacts $m_1$ to a much larger degree 
than $hk$, the agreement between the metallicities based upon the two 
indices is excellent.

\subsection{Comparison to Previous Determinations}
For the reddening determination, only one additional approach has become 
available since the discussion in \citet{tat97} where $E(B-V)$ = 0.06 was 
adopted. This estimate was defined by a number of approximate attempts to 
derive reddening from $UBV$ colors \citep{ha,vb} and crude constraints from 
isochrone fits \citep{bo,ca} that led to a range between $E(B-V)$ = 0.00 
and 0.08. \citet{ka} used a field RR Lyrae variable to place an upper limit 
of 0.08 on the cluster. The reddening maps of \citet{sc98} indicate 
$E(B-V)$ = 0.074 in the direction of NGC 2243, an upper limit along this 
line of sight but presumably close to the cluster value given the 
significant distance of the cluster above the galactic plane. In summary, 
given the uncertainty inherent in all of these techniques, there appears to 
be no significant discrepancy with the value of $E(B-V)$ = 0.055 derived 
from our $uvby$H$\beta$ photometry of the probable cluster members at the 
turnoff.

As with the reddening, the first attempts to derive a metallicity for NGC 
2243 were coupled to reddening estimation through analysis of $UBV$ 
photometry by \citet{ha} and by \citet{vb}. The former derived $E(B-V)$ = 
0.06 and $\delta(U-B)$ = +0.15 relative to the Hyades, while the latter 
found 0.03 and +0.06, respectively. Both indicated the cluster was 
metal-deficient, but the [Fe/H] values differed by 0.5 dex. The DDO data of 
\citet{noh} seemed to favor the unusually low [Fe/H] for an open cluster, a 
conclusion confirmed with the DDO recalibration \citep{ta96} and cluster 
reanalysis by \citet{tat97}, where DDO data from 5 giants produced [Fe/H] = 
--0.48 $\pm$ 0.16 (s.d.). Of the 5 stars included in this average, 2410 
lacks membership information, but 883 is assumed to be a nonmember based 
upon its anomalously high abundance \citep{fr02}. If this star is dropped 
from the
DDO sample, where its metallicity is also high, the mean [Fe/H] for the 
remaining
four stars becomes --0.53 $\pm$ 0.14(s.d.). The only additional photometric 
attempt
to measure the cluster metallicity uses the Washington system and, using 
the revised
calibration of \citet{ge}, implies [Fe/H] = --0.75 from 5 giants. It must 
be emphasized that
compared to the abundance scale of \citet{tat97} and \citet{fr02}, the 
Washington
[Fe/H] estimates for moderately metal-poor open clusters are generally too 
low by
between 0.1 and 0.3 dex. 

On the spectroscopic front, only two sources of [Fe/H] exist. High 
dispersion spectra 
of two giants analyzed by \citet{gr} give [Fe/H] = --0.48 $\pm$ 0.15, where 
the error
includes both internal and external uncertainties; the agreement between 
the two stars is much smaller. \citet{fr02} have expanded the original 
spectroscopic sample of \citet{fj} and revised the metallicity calibration 
based upon their spectra of the giants. The revision has compressed the 
scale by shifting the more metal-poor open clusters to higher [Fe/H] while 
lowering the metallicity of the clusters more metal-rich than the Hyades, 
exactly the trend delineated by comparison to the DDO data in 
\citet{tat97}. The result is that the earlier estimate for NGC 2243 of 
[Fe/H] = --0.56 from 6 stars \citep{fj} has improved to [Fe/H] = --0.49 
$\pm$ 0.05 from 9 stars. In summary, taking into consideration the 
differences among the techniques and the internal and external 
uncertainties in the calibrations and zero-points, there is every 
indication that the metallicity determinations for NGC 2243, whether based 
upon the evolved red giants or the turnoff stars, are converging toward a 
value of [Fe/H] = --0.55 $\pm$ 0.1 and, more importantly, systematically 
lower than the other well-studied anticenter open clusters, NGC 2204, NGC 
2420, NGC 2506, and Mel 66.

\section{Fundamental Parameters: Distance and Age}
\subsection{Defining the Giant Branch}
The traditional method for age and distance determination is via comparison 
of the cluster CMD to a well matched set of theoretical isochrones, the 
approach we have consistently used in this series. In previous papers, the 
cluster CMD was either richly populated, the cluster was nearby, and/or 
there was membership information available that allowed one to isolate the 
evolved members of the cluster from the field. Comparison of Figs. 3 and 5 
reveals that redward of $b-y$ = 0.45 and brighter than $V$ = 16.5, there is 
extensive contamination of the CMD by field stars along the line of sight, 
crossing the probable location of the subgiant branch between $V$ = 15.5 
and 16.5. One can minimize this issue by dealing solely with stars within 
the cluster core, a successful strategy for the easily identifiable turnoff 
and main sequence, but a concern for an entire post-main-sequence track 
defined by fewer than two dozen stars, some of which are likely to be field 
stars. Past
disagreement about potential contamination among the brighter stars 
\citep{vb,be}
can be tied to the limited coverage of earlier CCD surveys.  Questions 
about
the location of the red giant clump, as well as whether or not the cluster 
has more than
one clump, require as reliable an isolation of the members from the field 
as the data will allow, along with CCD-based photometry for all potential 
giants, not just those in the core \citep{bo}.
 
To define our giant branch, we will first limit our sample to all stars 
redder than $b-y$ = 0.45
and brighter than $V$ = 16.2; all stars in this portion of the CMD were 
transformed
photometrically as giants, with indices that smoothly transition to the 
dwarf calibration at
$b-y$ = 0.45. From this sample of 77 stars we eliminate 11 that have errors 
in $m_1$ 
and/or $hk$ greater than 0.015 mag, while retaining only stars with errors 
in $b-y$ below 0.010. One additional star that falls in the exclusionary 
range will be retained for reasons explained below. The strict selection 
criteria are necessary because we will be using the $m_1$ and $hk$ versus 
$b-y$ color-color diagrams to identify probable members. There is no proper 
motion data for the cluster and reliable radial velocities exist for 
approximately a dozen of the giants out of the 66 that remain. The sole 
advantage held by NGC 2243 is its almost uniquely low metallicity among the 
field stars in the disk. Since the cluster sits well above the galactic 
plane, most contamination is expected from field stars within the disk, 
stars that are expected to populate the two-color diagrams in the zone 
between Hyades metallicity and NGC2243. Few stars should be as metal-poor 
as NGC 2243.

To define the cluster relation in the two-color diagrams, we turn to the 
stars in the cluster
core, defined here as less than 300 pixels ($\sim$$1.5'$) from the cluster 
center. The boundary was chosen because it contains every star with a 
radial velocity consistent with 
cluster membership. The initial membership list is taken from the summary 
by \citet{fr02}. 
Of the 14 stars studied, 5 are classified as probable nonmembers. Of these 
stars, 3 are excluded due to their deviant velocities, while 2 have 
velocities consistent with 
membership, but spectroscopic abundances near solar. Two of the radial 
velocity 
members have been studied by \citet{gr} and confirmed as members. A tenth 
member 
based upon spectroscopic analysis (1654) has been added by \citet{hp}. An 
additional source of radial-velocity information is available from the 
moderate-dispersion spectroscopic survey of \citet{cr}. Unfortunately, the 
derived cluster velocity is systematically off from the true value by 15 
km/sec and the typical uncertainty in the velocity measures is close to 10 
km/sec. So, while stars near the cluster mean may or may not be members, we 
can use the data to exclude stars with exceptionally deviant velocities as 
probable nonmembers. To that list we add 2236 and 2704.

Armed with this information, one can now plot the $hk$ - $(b-y)$ diagram 
shown in Fig.
8 for stars within 300 pixels of the cluster core. Filled circles are 
cluster members, while
open circles are cluster nonmembers. With the exception of one star (2135), 
the members
form a tight sequence between $b-y$ = 0.58 and 0.9. The one exception is 
the star kept in the sample despite having unusually large scatter in its 
indices because it is a probable radial-velocity member. It is found at the 
level of the clump, redward of the first ascent giant branch and marked 
with a cross in Fig. 3. Its position is anomalously metal-poor relative to 
the cluster relation in both $hk$ and $m_1$. In earlier work, the star is 
identical to 3618 in $B-V$ \citep{ha}, bluer than 3618 by 0.06 mag in $V-I$ 
\citep{ka} , and redder than 3618 by 0.026 mag in $b-y$. The probable 
source of both the scatter among surveys and the internal scatter in $b-y$ 
is the crowding and contamination of a nearby star.

Assuming that the majority of the stars within the core are members and 
that they follow
a well-defined relation in the two-color diagram, we have extended the 
pattern defined
by the redder stars by identifying the starred points in Fig. 8 as members. 
Any star that
deviates from this pattern is classed as a nonmember (open squares). A 
similar figure was constructed for $m_1$ but is not shown; any star that 
deviated from the $m_1$ - $(b-y)$ relation was also classed as a nonmember. 
The agreement between the two diagrams for the core region is excellent. 
The members relation was then superposed upon the diagram for all 66 stars 
within the CCD survey and the same criterion applied. The more extensive 
scatter caused by field stars contributed by the outer zone is readily 
apparent in Fig. 9; the symbols have the same meaning as in Fig. 8. Note 
that a few points classed as non-members appear to fall within the 
restricted range defined by the members; these stars were excluded because 
of deviations in the $m_1 - (b-y)$ plot. The result is that by extending 
the analysis to stars outside the core, we add 6 highly probable members to 
the giant region, bringing the total for the survey to 27. 

The CMD for the stars in Fig. 9 is shown in Fig. 10 with the symbols having 
the same meaning. A number of points are apparent. First, a great deal of 
the scatter among the probable members in the CMD has been eliminated. The 
majority, though not all, of the stars that deviate from the first-ascent 
giant branch are classed as potential nonmembers while some of the stars 
that appear within the range of the giant branch are tagged as such. This 
is not unexpected given the photometric nature of the procedure and the 
likelihood that some of the remaining CMD deviants are composite members, 
especially with the rich binary population of the turnoff and the main 
sequence. As we have reiterated many times in this series, the exclusion of 
a few potential members from the sample is not a cause for concern as long 
as the stars that remain are highly probable cluster members.

Second, the identification of the red giant clump is now obvious. 
Seven probable members are found in the giant
branch region between $V$ = 13.63 and 13.81. Of these, the five bluer stars 
have an average $V$ of 13.70 $\pm$ 0.04 (s.d.) and an average $b-y$ of 
0.589 $\pm$ 0.007
(s.d.). In contrast, all the stars that populate the potential {\it fainter 
clump} between $V$
= 14.0 and 14.2 are classified as probable nonmembers. It should be noted, 
however,
that two of these stars (883, 2619) have radial velocities consistent with 
membership
but are excluded because they have spectroscopic abundances near solar 
\citep{fr02}.
Using our strict photometric criteria, both stars would have been excluded 
from the
sample anyway.

\subsection{The Broad-Band CMD - Transforming the Data}
As we have noted in past analyses, isochrones are invariably created for 
broad-band systems and a check of the most recent publications shows that  
theoretical isochrones are available on the $UBVRI$ system, among others, 
but rarely for $uvby$. This problem has been solved in past investigations 
by making use of the fact that $b-y$ is well correlated with $B-V$ at a 
given [Fe/H] with little dependence on evolutionary state.  While reliable 
CCD $BV$ data are available for the numerous main sequence stars in the 
cluster core, the
ability to transform from $b-y$ to $B-V$ is particularly important for the 
giants where more than half the identified probable members lie outside the 
published $BV$ surveys.

The first step is to merge the $BV$ photometry of BO and BE. For the $V$ 
magnitudes, both surveys have been placed on the system of Table 1 using 
the offsets and color terms of Table 2. For $B-V$, because of the more 
extensive tie-in of the CCD data of BE to the standard system, we have 
adopted BE as the color system of choice though, as noted below, choosing 
BO as the standard would have a negligible impact on our conclusions.  
Using only stars brighter than $V$ = 17.5 with a difference in $V$ between 
the two systems less than 0.075 mag, the difference in $B-V$, in the sense 
(BE-BO), has been calculated, producing a mean residual of 0.002 $\pm$ 
0.029 in $B-V$ from 190 stars. If we exclude 3 stars where the residuals in 
$B-V$ are larger than 0.10 mag, the average difference becomes 0.004 $\pm$ 
0.022. Clearly the two color systems are very similar. Finally, the 
residuals show a modest color term that can be removed by including a 
correction of $\Delta$$(B-V)$ = $0.072(B-V) - 0.035$; this reduces the 
residuals to 0.000 $\pm$ 0.019.  After applying the small color term to the 
data of BO, the composite system is constructed by taking a straight 
average for the modified $V$ and $B-V$ data, generating a final listing of 
$V, B-V$ data for over 900 stars. 

To transform the giants, we have selected the overlap between our sample of 
66 stars with $b-y$ $\geq$ 0.45 and $V$ $\leq$ 16.2 and the composite $BV$ 
data, resulting in 25 stars. Of these, three exhibit large deviations in 
the $(b-y)$ - $(B-V)$ transformation, 1804, 2410, and 3668. For the 
remaining 22 stars, the defined relation is $B-V$ = $1.647(b-y) - 0.045$; 
the scatter among the residuals about the mean relation is $\pm$0.022 mag. 
Of the three deviants, only 2410 is a probable member included in the 
discussion of the CMD as a red giant clump star, so its correct colors are 
important. Star 2410 was only observed in $B-V$ by BO. It has $B-V$ = 1.03, 
in contrast with clump stars 910, 1271, and 1467 at $B-V$ = 0.90, 0.91, and 
0.94, respectively. In the $VI$ survey of KA, the comparable colors are 
$V-I$ = 1.00 for 2410, and 0.97, 0.99, 0.99 for 910, 1271, and 1467. The 
latter pattern agrees well with $b-y$ where the four stars have $b-y$ = 
0.598, 0.581, 0.583, and 0.593. It appears that the modified $B-V$ color 
for 2410 from BO is systematically too large.  

For the remainder of the CMD, from the $V,B-V$ catalog we have selected 
stars from Table 1 located within 200 pixels of the cluster center with $V$ 
$\leq$18.5 for $B-V$ $\leq$ 0.8 and $V$ $\leq$ 16.2 for $b-y$ $\leq$ 0.45. 
If the differences in $V$ and/or $B-V$ between BE and BO were larger than 
0.15 mag or 0.10 mag, respectively, the star was excluded. The resulting 
CMD is shown in Fig. 11, where the giant probable members from the complete 
survey as discussed above have been drawn as filled circles; the cross 
signifies the anomalous star 2135. The CMD for the turnoff exhibits the 
striking features noted in the past CCD studies, though with slightly 
better delineation due to the combination of the photometry. The unevolved 
main sequence is easily identified down to $B-V$ = 0.65 and $V$ = 18.5, as 
is the rich, parallel band of binaries starting near $B-V$ = 0.8 and $V$ = 
18.5 that ultimately merges with the vertical turnoff near $V$ = 16.2. The 
reality of the small gap in the main sequence, discussed in every past 
study of the CMD, is undeniable and is enhanced by the fact that there is a 
small but real color shift on either side of the break. The stars above the 
gap are typically 0.01 to 0.02 mag bluer than the stars below the break. As 
will be discussed below, this small but significant feature plays a 
critical role in constraining the age and metallicity of the cluster.

\subsection{The CMD Fit: The Isochrones}
There is a variety of isochrone sets available for comparison with 
broad-band photometry. For consistency with our previous discussions and 
because, when properly zeroed to the same color and absolute magnitude 
scale, most sets produce similar results for ages and distances, we will 
use the sets of \citet{gl02} (hereinafter referred to as PAD). Moreover, 
based upon the many comparisons, including our own 
\citep{at91,da94,ash,tah}, between open clusters and past and present 
generations of isochrones, we will only make use of isochrones that include 
convective overshoot mixing. On a scale where solar metallicity is Z = 
0.019 and Y = 0.273, PAD isochrones were obtained for (Y, Z) = (0.25, 
0.008) and (0.24, 0.004) or [Fe/H] = --0.38 and --0.68, respectively. Since 
isochrones with [Fe/H] = --0.57 are not available, it was decided to define 
the cluster parameters assuming each of the abundances was appropriate and 
interpolate the results for the derived [Fe/H].  

For distance and age estimation the next step in comparing the cluster CMD 
to a set of theoretical isochrones is ensuring that the color 
transformations and bolometric corrections 
between the theoretical and the observational plane reproduce the colors 
and absolute magnitudes of nearby stars with known temperatures and 
abundances. From an age standpoint, the critical test is whether or not a 
star of solar mass, composition, and age resembles the sun. For open 
clusters of solar and subsolar abundance, the impact of this issue on 
cluster ages and distances has been emphasized on a variety of occasions 
\citep{tat89,tam,da94,tah,tab}. In past uses of the isochrones for clusters 
with approximately solar metallicity, we have checked to ensure that the 
isochrones were on our adopted color and absolute magnitude scale of $M_V$ 
= 4.84 and $B-V$ = 0.65 for a solar mass star at 4.6 Gyrs. One quickly 
finds that the solar metallicity isochrones of PAD are too red by 0.032 mag 
in $B-V$ and too bright by 0.02 mag in $V$; for the analyses of IC 4651 and 
NGC 3680, these adjustments were included. 

Should these offsets be applied to isochrones that are metal-deficient by a 
factor of five compared to the sun? There is some evidence that the 
correction for linking the metal-poor isochrones to the observational plane 
is not only smaller, it may be zero for the PAD isochrones \citep{tab}. 
While this implies that the distance modulus derived by comparing the 
cluster CMD to the unevolved, uncorrected main sequence of the isochrones 
should be correct, there is no independent way of zeroing the age scale as 
there is for solar metallicity. We will use the metal-deficient isochrones 
without adjustment, keeping in mind that while relative comparisons among 
the isochrones should be reliable, the absolute age and distance scales are 
subject to potential systematic offsets. 

The first comparison between theory and observation is for the [Fe/H] = 
--0.38 isochrones with ages 2.51, 2.82 and 3.16 Gyr, shown in Fig. 12 with 
a shift of $E(B-V)$ = 0.055 and an apparent distance modulus of $(m-M)$ = 
13.40. The color of the turnoff and the location of the subgiant branch are 
reasonably consistent with an age of about 2.7 Gyr, keeping in mind the 
nonlinear rate of evolution of both of these features. The isochrones, 
however, fail to match the cluster morphology in a number of ways. While 
the break in the turnoff due to hydrogen exhaustion has the correct 
approximate apparent magnitude, the predicted curvature of the turnoff 
below the break and the blueward extension of the hook above the break are 
too large. Such a small hook and break 
is predicted for a significantly older isochrone and is commonly associated 
with clusters similar in age to the sun such as M67. The more dramatic 
discrepancy is the location of the giant branch. The isochrone tracks are 
too red by 0.1 mag in $(B-V)$ and the predicted clump is too faint by 0.3 
mag. Note that with this modulus, the observed clump has $M_V$ = +0.3, 
making it anomalously bright compared to other clusters of similar age and 
metallicity \citep{tat97}.

In Fig. 13, we illustrate the second comparison with isochrones of [Fe/H] = 
--0.68; the reddening adjustment is the same, but the apparent modulus has 
been reduced to 13.05 to match the main sequence. The isochrones have ages 
of 3.16, 4, and 5 Gyrs. Based upon the color of the turnoff, the age of NGC 
2243 rises to 5 Gyr, though the position of the subgiant branch implies a 
value between 4 and 4.5 Gyr. Morphologically, the isochrones do a better 
job with the hydrogen exhaustion phase which is predicted to disappear 
between 4 and 5 Gyr, though the observed luminosity of the break is too 
high. The dramatic improvement occurs in the color of the giant branch and 
the red giant clump, where the isochrones are too blue by a few hundredths 
in $B-V$, and in the luminosity of the clump. The observed clump is too 
faint compared to the predicted position by 0.1 mag. With this distance 
modulus, the observed clump has $M_V$ = 0.65, much more consistent with the 
typical range between 0.5 and 0.7 for clusters younger than 5 Gyr 
\citep{tat97}.

The comparisons are exactly what one would expect for a cluster whose 
metallicity is intermediate between [Fe/H] = --0.38 and --0.68. Adopting 
[Fe/H] = --0.57, NGC 2243 is approximately one-third of the way between the 
limiting models. Interpolating the disparities noted above, isochrones with 
an appropriate [Fe/H] and adjusted for an apparent modulus of 13.15 should 
simultaneously match the color and luminosity of the giant branch and the 
clump, implying $M_V$ for the clump of +0.55. The age of the cluster is 
somewhat more of a challenge since the subgiant branch and the turnoff do 
not supply identical ages for the lower metallicity comparison. 
Approximating 4.6 Gyr as the age from [Fe/H] = --0.68, and interpolating in 
log(age) to account for the nonlinear change in age with [Fe/H] leads to 
3.8 $\pm$ 0.2 Gyr as the approximate age of NGC 2243. 

\subsection{Comparisons to Previous Work}
Past derivations of the cluster age and distance using CCD photometry are 
somewhat
dated because of the combination of adopted isochrones and cluster 
parameters, but the analyses do illustrate the convergence that has 
occurred in defining cluster parameters over the last decade as the range 
of models and their transformation to the observational plane have 
improved.

BE used a comparison of NGC 2243 to 47 Tuc and theoretical isochrones 
without
convective overshoot to conclude that NGC 2243 was likely to be more 
metal-rich than the globular cluster, in contradiction with some 
observational evidence at the time that NGC 2243 might be as 
metal-deficient as 47 Tuc, if not more so. Assuming similar abundances for 
the two clusters, BE derived $(m-M)$ = 13.05 for NGC 2243, with an age 
range of 4 to 6 Gyr for an [Fe/H] range of --0.47 to --0.78. It was 
concluded that convective overshoot was required to reproduce the main 
sequence gap, a correct interpretation, while oxygen-enhancement was needed 
to reproduce the color of the giant branch, a now questionable 
interpretation.

BO used cluster CMD morphology and synthetic CMDs based upon a variety of
theoretical stellar models in an attempt to constrain the cluster 
parameters. The models ranged from those with no convective overshoot, 
similar to what was adopted in BE, to extreme amounts of overshoot. With 
the exception of the extreme overshoot case where the main sequence gap 
could not be reproduced under any circumstances, BO were unable to decide 
which among the various evolutionary codes achieved a better match to the 
cluster CMD. Despite the ambiguity, BO concluded that the cluster must have 
Z between 0.003 and 0.006, $E(B-V)$ between 0.06 and 0.08, an age between 3 
and 5 Gyr, and a true modulus between 12.7 and 12.8 [$(m-M)$ between 12.9 
and 13.05], all values in surprisingly good agreement with what has been 
derived in the current investigation.

Finally, using the data of BO, $E(B-V) = 0.06$ and [Fe/H] = --0.44, 
\citet{tat97} found $(m-M)$ 
= 13.45, in good agreement with the result from the metal-rich isochrone 
match.

\subsection{Comparison to Berkeley 29}
With its slightly lower [Fe/H] determination and smaller distance modulus, 
NGC 2243 goes from being one of the three lowest [Fe/H] clusters in the 76 
cluster sample of \citet{tat97} to the lowest in the list, though the exact 
abundance for its nearest competitor, Berkeley 21, remains much more 
uncertain. The smaller distance to the cluster moves it closer to both the 
sun and the galactic center, reducing the galactocentric distance to 10.93 
kpc, on a scale where the sun is at 8.5 kpc, and just over the edge of the 
disk discontinuity. In the
revised and expanded sample of \citet{fr02}, only Ber 21 at [Fe/H] = --0.62 
and Ber 20 at [Fe/H] = --0.61 are below --0.57. Thus, there appears to be 
at least some overlap in [Fe/H] between the cluster population at a 
galactocentric distance of 11 kpc with that beyond 16 kpc, though the 
sample beyond 13 kpc is statistically challenged. 

This has led to an expanded interest in observing and defining the 
parameters for clusters at and beyond the 13 kpc barrier in the hopes of 
minimizing the impact of the small sample and greater abundance errors by 
greatly extending the distance baseline. A critical object for this mission 
is Ber 29, the subject of photometric study by \citet{pj, ka94} and, most 
recently, \citet{to}, and the claimant to the title of the most distant 
open cluster known, though exactly how far away it is depends greatly upon 
the adopted metallicity and reddening. Likewise, the cluster CMD morphology 
has been used to derive an age that is moderately old, between 2 and 5 Gyr. 

Given the age and/or CMD morphology, attempts have been made by various 
authors to infer a metallicity, again coupled to the reddening. 
\citet{ka94} used the reddening maps of \citet{bu} to assign the cluster an 
$E(B-V)$ greater than 0.2 mag. With an age close to M67 for the cluster, 
this required a very low metallicity, [Fe/H] below --1.0, to match the blue 
turnoff with the high age. \citet{nm} used a CMD morphology approach to 
simultaneously derive [Fe/H] = --0.30 and $E(B-V)$ = 0.01. While no single 
cluster should be used to define the abundance gradient, it is apparent 
than the conclusions one would draw are very different for each of these 
choices.

Fortunately, two recent studies have taken a more detailed look at the 
issue of the cluster age and abundance, \citet{gc04} and \citet{to}. The 
former discusses the results from high dispersion spectroscopy of 2 giants, 
while the latter uses CMD analysis and moderate dispersion data from 20 
giants, not all of which are members. \citet{gc04} find [Fe/H] = --0.44 
$\pm$0.18, while the spectroscopy of \citet{bra} produces [Fe/H] = --0.74 
$\pm$ 0.18. While statistically the two results overlap, as with past 
discussions of this cluster, the conclusions regarding the disk gradient 
can differ significantly depending upon which one is chosen.

Can our investigation of NGC 2243 in any way help clarify the uncertainty 
regarding this cluster within the abundance gradient? If one looks at the 
many morphological parameters that are available to define the age of Ber 
29, the consistent pattern that emerges from MAR \citep{att85,tat89}, 
$\delta$ \citep{jp94}, $\Delta$$V$ and $\Delta$$V_2$ \citep{ka94}, and 
$\delta$$(V)$ \citep{sa} is that Ber 29 is the same age as or slightly 
younger than NGC 2243. In the most recent determination by \citet{sa}, the 
difference in age is close to 0.5 Gyr after adjusting the [Fe/H] of Ber 29 
from their adopted value of --0.18 to a value near --0.5.
This simple parametric approach is confirmed by the synthetic CMD work of 
\citet{to}. Though the isochrones have changed since the study of NGC 2243 
by BO, \citet{to} reach virtually identical conclusions regarding the age 
and metallicity of Ber 29. The latter cluster should have [Fe/H] = --0.5 or 
--0.7, an age of 3.4 or 3.8 Gyr, with derived reddening of $E(B-V)$ = 0.13 
or 0.10.

The last critical piece of information which could have resolved the 
question with the original study of \citet{ka94} is the reddening. If one 
uses the reddening maps of \citet{sc98}, the reddening for the galactic 
field in the direction of Ber 29 is $E(B-V)$ = 0.093, on the same scale 
where the field of NGC 2243 has $E(B-V)$ = 0.074. This value is almost 
certainly the correct value for Ber 29 given its location almost 2 kpc 
above the galactic plane; if not, it should be an upper limit. 
Observational support for this claim comes from the spectroscopic analysis 
by \citet{gc04}, where the temperature scale defined by minimizing the 
slope of  abundances from the Fe I lines with respect to the excitation 
potential in the curve of growth analysis yields $E(B-V)$ = 0.08.

We will assume that NGC 2243 and Ber 29 have identical abundances. We can 
then superpose the CMD for Ber 29 on that of NGC 2243 by lowering the 
colors of Ber 29 by 0.038 mag in $B-V$ and shifting $V$ to account for the 
differential apparent modulus until the main sequences superpose, in this 
case 3.00 magnitudes, implying an apparent modulus for Ber 29 of 16.15 or a 
true modulus of $(m-M)_0$ = 15.85 (14.8 kpc), a little larger than the 
range of 15.6 to 15.8 predicted by \citet{to}. The data for Ber 29 have 
been taken from \citet{ka94} which fortunately is on the same system within 
a few millimagnitudes as the photometry of \citet{to}. To maximize the 
cluster membership, we will use only stars within 150 pixels of the cluster 
center as defined by \citet{ka94}; for NGC 2243 we will use the same stars 
found in Fig. 13. The result is shown in Fig. 14, where the points from NGC 
2243 are drawn in red. 

The agreement between the two clusters is impressive; the position of 
virtually every feature of the CMD is almost superposed for the two 
clusters, including the location of the main sequence gap. What little 
differences there are in the color of the turnoff, the extent of the blue 
hook, the relative location of the giants and the clump are easily 
reconciled by assuming that, morphologically, Ber 29 is slightly younger 
than NGC 2243, as predicted by the parameters discussed above. For a color 
difference of 0.02 in $B-V$ for the turnoff, in the age range between 3.2 
and 4 Gyrs, the age differential implied is approximately 0.4 Gyr, in 
excellent agreement with the prediction of \citet{sa}. What is important to 
emphasize is that this explanation can only work if Ber 29 is as metal-rich 
as NGC 2243. Any attempt to significantly lower the metallicity of Ber 29 
relative to NGC 2243 will result in an age that must be greater than that 
of NGC 2243, unless the reddening for Ber 29 is increased above the 
limiting value. Taking the age differential from \citet{sa} as our limit, 
if NGC 2243 has an age of 3.8 Gyr, Ber 29 must be 3.3 to 3.4 Gyr old, 
confirming the analysis of \citet{to}. All things being equal, on the scale 
of [Fe/H] = --0.57 for NGC 2243 and the isochrones adopted above as zeroed 
to the observational plane, with the intrinsic color of the turnoff, 
$(B-V)_0$, implied by a reddening of $E(B-V)$ = 0.093, [Fe/H] for Ber 29 
must be --0.57 or above.  

Not surprisingly, all things are not equal. One of the conclusions of the 
spectroscopic work of \citet{gc04} is that the giants of Ber 29 have 
non-solar abundances for some of the light alpha elements, {\it i.e.,} 
they have enhanced abundances, in contrast with NGC 2243 where the elements 
appear to be scaled solar. What impact does this have on the analysis?

A simple test of the impact can be made by a comparison of the CMD for Ber 
29 to a set of isochrones with the appropriate [Fe/H] but other elements 
not scaled to the sun. The lowest [Fe/H] value for which the alpha-enhanced 
isochrones are available from \citet{gl02} is --0.38, slightly higher than 
required, but still indicative of the effect of the changes on the 
morphology and the age of the cluster; these are the same models adopted by 
\citet{gc04}.
Fig. 15 shows the photometry of Ber 29 superposed upon isochrones shifted 
by $E(B-V)$ = 0.093 and an apparent modulus of $(m-M)$ = 16.25. The match 
of the data to the isochrones, as defined by the color of the turnoff and 
the luminosity of the subgiant branch, gives an age of 2.8 Gyr. Had we used 
the scaled solar abundance isochrones with [Fe/H] = --0.38 as we did for 
NGC 2243, the age for Ber 29 would be 2.5 Gyr, slightly younger than NGC 
2243 when compared to the same set, as expected. The change from the 
earlier comparison of NGC 2243 is the exceptional agreement with the color 
and luminosity of the giant branch and red giant clump. 
If we assume the age effect of boosting the alpha elements is approximately 
10\% in this age range, assuming that Ber 29 has the same [Fe/H] as NGC 
2243 raises its age to about 3.7 Gyr, the same value within the errors as 
NGC 2243. If we adopt [Fe/H] = --0.44 as measured by \citet{gc04}, the age 
must be lower; we estimate 3.1 Gyr. \citet{gc04} claim an age of 4.5 Gyr 
for Ber 29 using the same isochrones, but provide no details, so the source 
of the discrepancy remains unknown.

As an unrelated item, during the analysis of Ber 29 a check of the 
discussion by \citet{ka94} of the cluster Ber 54 identified a large error 
in the adopted distance modulus for this cluster. \citet{ka94} finds, via 
differential comparison with M67, a cluster with the same morphological 
age, that Ber 54 is heavily reddened at $E(B-V)$ = 0.77 but with a true 
distance modulus of $(m-M)_0$ = 11.8 or a distance of 2.3 kpc. Using these 
numbers, the apparent modulus is $(m-M)$ = 14.2. Since the cluster turnoff 
point is fainter than $V$ = 19.5, this makes no sense. The true distance 
modulus of 11.8 has propagated through the literature \citep{fr,sa}, 
placing the cluster at a galactocentric distance of 8.54 kpc on a scale 
where the sun is at 8.5 kpc. 

An alternative approach is to assume that the red giant clump at $V$ 
$\sim$17.3 has an absolute magnitude typical of clusters with an age near 
M67, $M_V$ = +0.6 \citep{tat97}.
Thus, the apparent modulus becomes $(m-M)$ = 16.7 and the true modulus is 
$(m-M)_0$ = 14.3. The similarity of this number with the conclusion of 
\citet{ka94} supplies the likely explanation that the apparent and true 
moduli were transposed. On our scale, Ber 54 has a distance of 7.2 kpc from 
the sun and a galactocentric distance of 10.5 kpc, well beyond the solar 
circle.

\section{Summary and Conclusions}
The specific goal of this investigation has been the derivation of the key 
cluster parameters of reddening, metallicity, distance, and age for the 
metal-poor anticenter cluster, NGC 2243.  We have attempted to use the 
various photometric indices to optimize the sample of single-star cluster 
members near the turnoff with high precision photometry in determining  
$E(B-V)$ = 0.055 $\pm$ 0.004 and [Fe/H] = --0.57 $\pm$ 0.03, with excellent 
agreement among the various combinations of indices used to derive [Fe/H]. 
This point is crucial because $m_1$, $hk$,  $b-y$, and H$\beta$ are tied to 
the standard system independently, {\it i.e.}, each has its own 
transformation from instrumental to standard system. If there were 
significant errors in the zero-points of the indices, the variation in the 
dependence of each color index on reddening and metallicity would conspire 
to generate much greater differences among the derived metallicities from 
the four techniques than are found in the analysis. In fact, the majority 
of the scatter among the derived [Fe/H] values can be explained solely by 
the internal scatter within the photometry for the various indices.

Given the reddening and metallicity, one should be able to obtain the 
distance and age by comparison to appropriate isochrones. Using color-color 
diagrams coupled with the modest amount of membership information supplied 
by high quality radial velocities, we were able to isolate highly probable 
members from the extended cluster area rather than just the core. This 
proved valuable in that it excluded a large number of field giants that 
confused the location of the subgiant branch and eliminated the 
controversial second clump from the discussion as 
a random projection of field stars. The note of caution that underlies this 
claim is that two of these stars have radial velocities consistent with 
membership, but are excluded because their abundances are inconsistent with 
those of the cluster. If some form of anomalous evolution has the ability 
to make normal cluster stars appear metal-rich because of mass transfer, 
mixing, or some unknown phenomenon, these stars should be revisited. 
However, for our purposes, their exclusion is reasonable.

Comparison of the cluster CMD to isochrones that bracket the derived [Fe/H] 
produces a set of parameters and discrepancies indicating that neither 
[Fe/H] is ideal, but that NGC 2243 lies at an [Fe/H] intermediate to that 
of the sets tested. The final, best fit parameters are $(m-M)$ = 13.15 
$\pm$0.10 at an age of 3.8 $\pm$0.2 Gyr. As always, it cannot be 
overemphasized that these estimates are tied to a specific set of 
isochrones generated using a specific conversion between the theoretical 
and observational plane. In a relative sense, the data confirm past 
analyses using cluster morphology that NGC 2243 is approximately the same 
age as M67 but younger than Mel 66. The data should be on a the same scale 
as the clusters analyzed in previous papers in this series.

Given our results for NGC 2243, the cluster is located at a galactocentric 
distance interior to 11 kpc, but at the edge of the galactic discontinuity. 
A differential comparison between NGC 2243 and Ber 29 has been undertaken 
because the latter cluster now represents the most distant open cluster 
with a supposedly reliable [Fe/H] based upon spectroscopy. Unfortunately, 
the spectroscopic abundances differ by 0.3 dex, despite having overlapping 
one-sigma error bars, making the interpretation of the distant galactic 
gradient an exercise in personal bias. If one believes in a uniform 
gradient over a distance of 17 kpc, the high [Fe/H], when coupled with 
Saurer 1, forces the beholder to dismiss the clusters as anomalies 
unrepresentative of the real disk \citep{gc04}.
Adoption of the low metallicity is consistent with the claim of a linear 
galactic gradient 
\citep{to}, but leaves no explanation for Saurer 1. An even simpler 
solution that is consistent with the higher metallicity for Ber 29 and 
Saurer 1 is that there is no significant gradient beyond 10.5 kpc, as found 
by \citet{tat97}. The current reality is that if these two cluster have 
abundances typical of clusters near 11 kpc, it is intriguing, but hardly 
definitive evidence for any claim regarding the nature of the disk at 
galactocentric distances near 20 kpc. More and better data are needed both 
at large distance and in the key region bracketing 10 kpc.

Given the similar morphology of NGC 2243 and Ber 29, we can place some 
constraints on the differential properties of the two clusters  
due to the significant improvement in reddening estimation for Ber 29.
The reddening maps of \citet{sc98} coupled with the spectroscopic data of 
\citet{gc04} imply that Ber 29 has $E(B-V)$ below 0.094. If we adopt this 
limit as the true value for the cluster and assume that  Ber 29 has [Fe/H] 
identical to NGC 2243, a differential comparison of the two clusters' CMDs 
implies $(m-M)$ = 16.15 for Ber 29 and an age slightly younger than NGC 
2243. Any attempt to make Ber 29 significantly more metal-poor than NGC 
2243 would require that Ber 29 be older than NGC 2243 and/or that the 
reddening estimate for Ber 29 be raised above the limit. The derived 
parameters for Ber 29 are in surprisingly good agreement with those of 
\citet{to} and, when coupled to the success of the early analysis of NGC 
2243 by \citet{bo} in comparison with the current investigation, the 
convergence of the results is growing evidence of the power of the 
technique of CMD synthesis in generating reliable results in the absence of 
more direct observational data. Adjusting the age estimate for Ber 29 if 
[Fe/H] = --0.44 leads to 2.8 Gyr.

However, spectroscopic data \citep{gc04} indicate that Ber 29 does not have 
scaled-solar abundances and, unlike NGC 2243, has some light-alpha-enhanced 
abundances. If we choose to compare the cluster to the available isochrones 
with the closest match to this effect, the age estimates increase by 
approximately 10\% but, more importantly, the agreement between the 
location and luminosity of the giant branch and clump for [Fe/H] = --0.38 
is excellent, in sharp contrast with the results for NGC 2243. The apparent 
modulus for Ber 29 rises slightly to $(m-M)$ = 16.25. On the basis of the 
internal consistency of the comparisons made above, it would be difficult 
to assign an [Fe/H] as low as --0.8 to Ber 29; all of our results indicate 
that the value of --0.5 $\pm$0.1 is the most probable. 

\acknowledgements
The progress in this project would not have been possible without the time 
made available by the TAC and the invariably excellent support provided by 
the staff at CTIO.  We wish to thank the referee for thoughtful comments
that enabled us to improve the paper. 
Extensive use was made of the SIMBAD database, 
operating at CDS, Strasbourg, France  and the WEBDA database maintained at 
the University of Geneva, Switzerland. The cluster project has been helped 
by support supplied through the General Research Fund of the University of 
Kansas and from the Department of Physics and Astronomy.

\clearpage

\figcaption[fig1.eps]{Standard errors of the mean (sem) for the $V$, $b-y$, 
and $m_1$as a function of $V$. Open symbols are the average sem while the 
error bars denote the one sigma dispersion. Filled symbols are stars 
identified as potential variables. \label{fig1}} 

\figcaption[fig2.eps]{Same as Fig. 1 for the $c_1$, $hk$, and H$\beta$. 
\label{fig2}}

\figcaption[fig3.eps]{Color-magnitude diagram for stars with at least 2 
observations each in $b$ and $y$. Crosses are stars with internal errors in 
$b-y$ greater than 0.010 mag.\label{fig3}}

\figcaption[fig4.eps]{Surface density of stars as a function of radial 
distance from the cluster center for stars brighter than $V$ = 18.0 from 
Table 1 (circles), the radial trend from \citet{vb} (squares), and the 
radial trend from \citet{ka} (triangles). Curves have been normalized so 
that density is the same in the outer regions for all three samples. 
\label{fig4}}

\figcaption[fig5.eps]{Same as Fig. 3 for stars within 200 pixels of the 
cluster center. \label{fig5}}

\figcaption[fig6.eps] {$V,v-y$ CMD for cluster core stars at the turnoff. 
Filled circles are stars tagged as potential binaries, nonmembers, or 
photometric anomalies. \label{fig6}}

\figcaption[fig7.eps]{Traditional CMD for the same stars as Fig. 
6.\label{fig7}}

\figcaption[fig8.eps]{Two-color diagram for giants within 300 pixels of the 
cluster
center. Filled circles are known cluster members while open circles are 
nonmembers. Based upon the position in the diagram, squares are stars 
tagged as probable 
nonmembers while asterisks identify probable members. \label{fig8}}

\figcaption[fig9.eps]{Same as Fig. 8 for the entire CCD field. 
\label{fig9}}

\figcaption[fig10.eps]{CMD for the stars in Fig. 9. Symbols have the same 
meaning as in Fig. 8. \label{fig10}}

\figcaption[fig11.eps]{Broad-band CMD for NGC 2243 using stars from the 
cluster core
(open circles) and the transformed $b-y$ for the member stars in Fig. 9 
(filled circles). 
The cross symbol denotes star 2135. \label{fig11}}

\figcaption[fig12.eps]{Comparison of the PAD isochrones for [Fe/H] = --0.38 
with the
CMD for NGC 2243. The isochrones, ages 2.51, 2.82, and 3.16 Gyrs, have been 
adjusted
for reddening of $E(B-V)$ = 0.055 and an apparent modulus of $(m-M)$ = 
13.40. 
\label{fig12}}

\figcaption[fig13.eps]{Same as Fig. 12 using isochrones with ages of 3.16, 
4, and 5
Gyrs, [Fe/H] = --0.68, $E(B-V)$ = 0.055, and $(m-M)$ = 13.05. 
\label{fig13}}

\figcaption[fig14.eps]{CMD for the core of Ber 29 (open circles) adjusted 
by $\Delta(B-V) = -0.038$ and
$\Delta V = -3.00$ and superposed on the CMD for NGC 2243 (red crosses). 
\label{fig14} }

\figcaption[fig15.eps]{Comparison of the CMD of Berk 29 to alpha-enhanced 
isochrones with [Fe/H] = --0.38, adjusted for E$(B-V) = 0.093$ and 
$(m-M)=16.25$.  The isochones have ages of 2.51, 2.82 and 3.16 
Gyr. \label{fig15}}

\enddocument